\documentclass{ws-procs9x6}
\input symbols
\begin{document}

\title{Measurement of the CKM angles $\alpha$ and $\gamma$ at the $\babar$ experiment}

\author{Vincenzo Lombardo}

\address{\`Ecole Polytechnique, Laboratoire Leprince-Ringuet\\
F-91128 Palaiseau Cedex (France)\\
\vspace{0.2cm}
E-mail: lombardo@slac.stanford.edu}
\maketitle
{\small BABAR-PROC-07/003, SLAC-PUB-12530\\}
\abstracts{We present recent measurements of the CKM angles $\alpha$ and $\gamma$ using data collected 
by the \babar\ detector at the PEP-II asymmetric-energy $e^{+}e^{-}$ collider at the Stanford Linear 
Accelerator Center. In addition to constraints on $\alpha$ from the decays $B^0 \rightarrow \pi^+\pi^-$, 
$B^0\rightarrow\rho^{\pm}\pi^{\mp}$, and $B^0\rightarrow\rho^+ \rho^-$, we also report the first measurement 
of time-dependent \CP\ asymmetries in the decay $B^0 \rightarrow a_{1}^{\pm}(1260)\pi^{\mp}$.  We present
 measurements of $\gamma$ in $B^{\pm}\rightarrow D^{(*)0} K^{\pm}$ decays using a Dalitz analysis in the 
modes $D^{0}\rightarrow K_{s} \pi^+ \pi^-$ and $D^{0}\rightarrow\pi^+\pi^-\pi^0$.}

\section{Introduction}
\label{sec:intro}

The measurements of the angles $\alpha$, $\beta$ and $\gamma$ of the Unitarity Triangle (UT)
at the B-factories are providing precision tests of the description of \CP\ violation in 
the Standard Model (SM). This description is provided by the Cabibbo-Kobayashi-Maskawa (CKM) 
quark-mixing matrix\cite{CKM,wolfenstein}. I am summarizing here the experimental 
constraints on the Unitarity Triangle angle $\alpha$ and $\gamma$ obtained from \B\-meson decays 
with the \babar\ experiment at SLAC. The \babar\ detector and PEP-II accelerator are described 
elsewhere\cite{babar}.

\section{Measurements of the angle  $\alpha$}
\label{sec:alpha}

The decays of neutral \B\ mesons to the final states $hh'$, where $h^+,h'^-$=$\pi$, $\rho$, $a_1$ 
are sensitive to the CKM angle $\alpha$ in the interference between decay and mixing~\cite{a1}. 
 The presence of gluonic loop (``penguin'') contributions with a different weak phase to the tree 
contribution shifts the measured angle from the UT angle $\alpha$ to an effective parameter $\alpha_{eff}$, 
where the shift is defined as  $\delta \alpha = \alpha - \alpha_{eff}$. The time-dependent \CP\ asymmetry has the form:
\begin{equation} 
\mathbf{A(t)}=S\sin(\Delta m_d\Delta t)-C\cos(\Delta m_d\Delta t)
\end{equation}
where $\Delta m_d$ is the $B\bar B$ mixing frequency, $\Delta t$ is the proper time difference between 
the decay of the two \B\ mesons in an event and the coefficients are given by:
\begin{equation}
S=\frac{2{\rm Im}(\lambda)}{1+|\lambda|^2},\quad
C=\frac{1-|\lambda|^2}{1+|\lambda|^2}, \quad \lambda=\frac{q}{p}\frac{\bar{A}}{A}
=e^{2i\alpha}\frac{1-\frac{P}{T}e^{-i\alpha}}{1-\frac{P}{T}e^{+i\alpha}}
=|\lambda|e^{2i\alpha_{\rm eff}}
\label{eq:coeffs}
\end{equation}
where $q$ and $p$ are the \B\ mixing coefficients and $\frac{P}{T}$ is the
penguin to tree amplitude ratio, which can be different for $\pi\pi$,
$\rho\pi$, $\rho\rho$ and $a_1\pi$. Either isospin symmetry \cite{GL,Gross} 
or broken SU(3) flavor symmetry \cite{GZ} can be employed to disentangle $\alpha$ 
from $\alpha_{eff}$. 

\subsection{$\B\to\pi\pi$ and $\B\to\rho\rho$}
\label{subsec:alpha-pipi}

The measurements of the various branching fractions and \CP\ asymmetries measured in $\B\to\pi\pi$ 
and $\B\to\rho\rho$ are summarized in Tab.~\ref{tab:tablea}, ${A}_{\CP}$ is the charge (tag) asymmetry 
in the case of a charged
(neutral) \B\ decay.  The measurements are sufficiently well established
to perform an isospin analysis.  However, the value of $\BR(\B\to\piz\piz)$ is
the limiting factor in the $\B\to\pi\pi$ isospin analysis; its value is too large to allow a tight bound 
to be placed on $\delta\alpha$~\cite{pippim}. The present measurement excludes the absence of \CP\-violation 
($S_{\pi\pi}=0$, $C_{\pi\pi}=0$) at a C.L. of 3.6 $\sigma$. The limit that results from the current isospin 
analysis is $\delta\alpha^{\pi\pi}<41\degrees$ at 90\% C.L\cite{pippim}.

\begin{table}[ph]
\tbl{Summary of \babar\ measurements of $\B\to\pi\pi$ and $\B\to\rho\rho$ decays.}
{\footnotesize
\begin{tabular}{@{}|rccc|@{}}
\hline
 \hline
    Mode &  $\BR(10^{-6})$ & $S$ & $C$\\[1ex]
    \hline
    \pipi & $4.7\pm0.6\pm0.2$ & $-0.53\pm0.14\pm0.02$ & $-0.16\pm0.11\pm0.03$ \\[1ex]
    $\rho^+\rho^-$ & $23.5\pm2.2\pm4.1$ & $-0.19\pm0.2^{\,+\,0.05}_{\,-\,0.07}$ & $-0.07\pm0.15 \pm 0.06$ \\[1ex]
    \hline
    &  &  \multicolumn{2}{c|}{${A}_{\CP}$}\\[1ex]
    \hline
    $\rho^\pm\rho^0$ & $16.8\pm2.2\pm2.3$  & \multicolumn{2}{c|}{$-0.12\pm0.13\pm0.10$} \\[1ex]
    $\rho^0\rho^0$ & $1.07\pm0.33\pm0.19$  & \multicolumn{2}{c|}{---}\\[1ex]
    \hline
    $\pipm\piz$ & $5.12\pm0.47\pm0.29$ & \multicolumn{2}{c|}{$-0.01\pm0.10\pm0.02$}\\[1ex]
    $\piz\piz$ & $1.48\pm0.26\pm0.12$ & \multicolumn{2}{c|}{$-0.33\pm0.36\pm0.08$}\\[1ex]
    \hline
\end{tabular}
\label{tab:tablea} }
\vspace*{-13pt}
\end{table}
\noindent
The analysis of $\B\to\rho\rho$ is potentially complicated due to the
possible presence of three helicity states for the decay.  The
helicity zero state, which corresponds to longitudinal polarization of the
decay, is \CP-even but the helicity $\pm1$ states are not \CP\ eigenstates.
Fortunately this complication is avoided due to the experimental determination
that the longitudinally polarized fraction is dominant $f_L=0.977\pm0.024(stat)^{+0.015}_{-0.013}(syst)$. 
This and other $\rho\rho$
measurements are summarised in ~Tab.~\ref{tab:tablea}.
The measurements of the branching fractions of $\B\to\rho^\pm\rho^0$ and
$\B\to\rho^0\rho^0$ indicate that the penguin pollution is small in these
modes compared with $\B\to\pi\pi$ decays~\cite{rhozrhop}~\cite{rhozrhoz}.  As such it is possible to perform an
isospin analysis on the longitudinal part of the decay and to place a much
tighter bound on $\delta\alpha^{\rho\rho}$;  the measured \CP\-violating parameters 
in $\B\to\rho^+\rho^-$ corresponds to $\alpha_{eff}^{\rho\rho}$ = $(95.5^{~+6.9}_{~-6.2})\degrees$ 
and the limit that results from the current isospin analysis is $\delta\alpha^{\rho\rho}<20\degrees$ 
at 90\% confidence level (C.L.)\cite{rhozrhoz}. 

\subsection{$\B\to\rho\pi$ and $\B\to a_1\pi$}
\label{subsec:alpha-rhopi}

The $\B\to\rho\pi$ measurement reported here is a time-dependent Dalitz plot analysis. We model the
interference between the intersecting $\rho$ resonance bands and so determins
the strong phase differences from the Dalitz plot structure\cite{snyder-quinn}.
The Dalitz amplitudes and time-dependence are contained in complex
parameters that are determined by a likelihood fit. The values obtained for these
parameters are then converted back into the quasi-two-body \CP\ observables, $S$, $C$, $\Delta S$, 
$\Delta C$ and $A_{CP}$ which are more intuitive in their interpretation\cite{rhopi-q2b}.

\begin{table}[ph]
\tbl{Summary of the \babar\ quasi-two-body \CP\ observables in $\B\to\rho\pi$ and $\B\to a_1\pi$ decays. 
The parameters $\Delta S$ and $\Delta C$ are insensitive to \CP\ violation.}
{\footnotesize
\begin{tabular}{@{}|rccc|@{}}
  \hline
  \hline
  Mode &  $S$ & $C$& ${A}_{\CP}$\\ [1ex]
      \hline
 $\rho^{\pm}\pi^{\mp}$ &$0.010\pm0.120\pm0.028$ &$0.154\pm0.090\pm0.037$ & $-0.142\pm0.041\pm0.015$ \\[1ex]
      $a_1^{\pm}\pi^{\mp}$ &$0.37\pm0.21\pm0.07$ &$-0.10\pm0.15\pm0.09$ & $-0.07\pm0.07\pm0.02$ \\[1ex]
      \hline
      &$\Delta S$& $\Delta C$&\\[1ex]
      \hline
      $\rho^{\pm}\pi^{\mp}$  &$0.060\pm0.130\pm0.029$&$0.377\pm0.091\pm0.021$& \\[1ex]
      $a_1^{\pm}\pi^{\mp}$&$-0.14\pm0.21\pm0.06$&$0.26\pm0.15\pm0.07$ &\\[1ex]
      \hline
\end{tabular}\label{tab:table2} }
\vspace*{-10pt}
\end{table}
\noindent
Using these results we obtain $\alpha^{\rho\pi} \in$ (75,152)\degrees at 68\% C.L. This result is of particular 
interest because there is a unique solution between
0 and 180\degrees, which helps to break the ambiguity on the $\rho\rho$
result, which is in itself more precise. We get a hint of direct \CP\ violation at the $3.0~\sigma$ level. \\
The first measurements of \CP-violating asymmetries in $\B\to a_1\pi$ decays with 
$a_1^{\pm} \rightarrow \pi^{\pm}\pi^{\mp}\pi^{\pm}$ have recently been performed by \babar\ 
using a ``quasi-two-body''approach\cite{lombardo}. A full isospin analysis requires the precise 
measurements of the branching fractions and asymmetries in the five modes 
$B^0 \rightarrow a_1 ^+\pi^-, a_1 ^-\pi^+, a_1^0\pi^0, B^+ \rightarrow a_1^+ \pi^0, a_1^0 \pi^+$ 
and in the charged conjugate modes. However, even measuring all the branching fractions and 
time-dependent \CP asymmetries in the three $B^0$ decay modes, this isospin method for extracting 
the angle $\alpha$ is not feasible with the present statistics. Assuming flavor SU(3) symmetry one 
can determine an upper bound on $\delta \alpha^{a_1\pi}$ using SU(3) related decays to $a_1 \pi$~\cite{zupan}. 
The measured \CP parameters in this mode are shown in Tab.~\ref{tab:table2}. Using these quatities 
$\alpha_{eff}^{a_1\pi}$ = $(78.6\pm7.3)\degrees$ has been extracted~\cite{lombardo}. Once the measurements 
of branching fractions for the SU(3)-related decays become available, an upper bound on $\delta \alpha^{a_1\pi}$ 
will provide a constraint on the angle $\alpha$.   

\section{Measurements of the angle  $\gamma$}
\label{sec:gamma}

Sensitivity to the CKM angle $\gamma$ occurs in decay modes that have
contributions from diagrams containing $\b\to\c$ and $\b\to\u$ transitions
that interfere with eachother.  The size of the interference, and hence the
sensitivity to $\gamma$, is determined by the relative magnitudes of the two
processes.  The two diagrams considered here are those of
$\Bp\to\Dzb\Kp$ and $\Bp\to\Dz\Kp$. In order for these two processes to 
interfere it is required that the final state be the same. Here we examine the decay of the \Dz\ and \Dzb\ to $\KS\pipi$. 
In this decay mode, there are four unknowns $\gamma$, $r_\B\equiv\frac{|A(\Bp\to\Dz\Kp)|}{|A(\Bp\to\Dzb\Kp)|}$, $\delta_\B$ 
(the strong phase of the \B\ decay) and $\delta_D$ (the strong phase of the $D$ decay). This last parameter is eliminated
 by using the Dalitz plot structure of the $\Dz\to\KS\pipi$ decay in the likelihood fit.  This is determined by performing 
a full Dalitz plot analysis of this $D$ decay mode using a very high statistics sample of \Dstarp\ decays.  The resulting 
amplitude model is
then fixed in the fit. A simultaneous fit is then performed to the \Bp\ and \Bm\ data samples in order to determine $\gamma$, 
$\delta_\B$ and $r_\B$. In addition to the Dalitz plot information, kinematic and event topology information is used to 
separate the signal and background events~\cite{gamma}. We obtain $\gamma=(92\pm41(stat)\pm11(syst)\pm12(theo))$\degrees. 
Preliminary results in $\Bm\to\Dz\Km$ decays with $\Dz\to K^-\pi^+\pi^0$ and $\Dz\to \pi^+\pi^-\pi^0$ have been presented, 
their effect on $\gamma$ have not been evaluated yet~\cite{gamma1,gamma2}.

\section{Summary}
\label{sec:summary}
The \babar\ experiment has conducted several analyses with the aim of extracting $\alpha$ and $\gamma$. In the last few 
years the measurements of the angles of the CKM Unitarity Triangle have become increasingly sophisticated and precise. 
At present the \babar\ measurement of the alpha and gamma angles are in a good agreement with the predictions obtained 
by SM-based fits.

\section*{Acknowledgements}
Many thanks to Fernando Palombo for helpful discussions. Some measurements presented at Lake Louise Winter Institute 
2007 have recently been updated by \babar\ . The direct \CP\ asymmetry has been observed in $\B^0\to\pi^+\pi^-$ 
decays~\cite{telnov}. $\B^0\to\rho^{\pm}\pi^{\mp}$ and $\B^0\to\rho^+\rho^-$ analyses have been updated using the 
full \babar\ dataset~\cite{matt,adrian}.


\begin{thebibliography}{0}
\bibitem{a1}
For the $a_1(1260)$ meson we use the short notation ${a_1}$.
\end{thebibliography}
\end{document}